\def\Mn12{Mn$_{12}ac$}
\def\cm{{\rm cm}$^{-1}$}
\documentstyle[prl,aps,graphicx,multicol,amsmath,amssymb]{revtex}

\begin{document}
\draft
\title{Quantum Tunneling and Relaxation in Mn$_{12}ac$
Studied by Magnetic Spectroscopy}
\author{M. Dressel$^1$\cite{email}, B. Gorshunov$^{1,2}$,
K. Rajagopal$^1$, S. Vongtragool$^1$, and A.A. Mukhin$^{1,2}$}
\address{$^1$ 1.\ Physikalisches Institut, Universit\"at Stuttgart,
Pfaffenwaldring 57, D-70550 Stuttgart, Germany\\
$^2$ General Physics Institute, Russian Academy of Sciences, 38
Vavilov St., 119991 Moscow, Russia}
\date{Received  \today}
\maketitle

\begin{abstract}
The spin relaxation in the molecular magnet \Mn12\ is investigated by a
novel type of high-frequency magnetic spectroscopy. By reversing
the external magnetic field the intensities of the
$|\pm 10\rangle\rightarrow |\pm 9\rangle$ transitions within the ground multiplet redistribute;
below $T\approx 2$~K resonant quantum tunneling is directly observed as
{\em magnetic hole burning} in the absorption spectra of a polycrystal. We
quantitatively describe our observations by taking into account inhomogeneous
line broadening, the time dependence of the level population
during the relaxation process, and the resonance behavior of the
relaxation rate due to quantum tunneling near the avoided
level crossing.
\end{abstract}


\begin{multicols}{2}
\columnseprule 0pt 
\narrowtext 
In the last years molecular magnets
comprising large clusters of coupled magnetic ions have attracted
considerable interest as mesoscopic systems which exhibit  new
properties such as molecular magnetic bistability, macroscopic
quantum tunneling of magnetization, quantum phase interference,
etc.\ \cite{Gatteschi94,Friedman96,Barbara99b,Lionti97}. Among them clusters with a high-spin ground
state and large anisotropy,
such as Mn$_{12}$O$_{12}$(CH$_{3}$COO)$_{16}$(H$_{2}$O)$_{4}$]2CH$_{3}$COOH%
$\cdot$4H$_{2}$O (ab\-bre\-vi\-ated as \Mn12) are of a paramount
interest.

\Mn12\ cluster can be considered as nanoparticle with an effective
magnetic moment of 20$\mu_B$ ($\mu_B$ is the Bohr magneton),
corresponding to a collective spin $S=10$ of the exchange coupled
twelve Mn ions within the molecule. Due to the strong anisotropy,
the magnetic moment freezes along one of the two easy directions
at low temperatures. This is governed by a crystal field (CF)
which splits the ground $S=10$ multiplet and provides a
quasidoublet structure $|\pm m\rangle$ of the lowest energy
levels. The energy barrier $\Delta E\approx 60$~K between the two
lowest lying states $|\pm 10\rangle$ determines the thermally
activated relaxation of the magnetization in \Mn12\
\cite{Barbara99b,Lionti97}. Below a blocking temperature
of about 3~K resonant quantum tunneling was discovered in the form
of peaks in the relaxation rate and steps in the hysteresis loops
at regular intervals of magnetic field corresponding to a
coincidence of the CF energy levels \cite{Gatteschi94,Friedman96}.
Since the rate of the tunneling for a ground state is small in
\Mn12\, the tunneling occurs via appropriate thermally excited CF
states for temperatures above approximately 2 K resulting in a
shortcut at the top of the barrier which accelerates the
relaxation; at even lower temperatures pure
quantum tunneling is dominant \cite{Barbara99b,Lionti97}.

Various spectroscopic investigations have been applied to \Mn12,
including electron-spin resonance (ESR) \cite{Hill97,Barra97} and
inelastic neutron scattering \cite{Hennion97}. In particular, a
new type of high-frequency ESR method based on a quasi-optical
spectroscopy \cite{Mukhin98} made it possible to directly observe
the CF transitions in an equilibrium state and to study in details
their characteristics, including the lineshapes \cite{Mukhin01}.
In this Letter we report on the high-frequency magnetic
spectroscopy of {\em non-equilibrium} phenomena in \Mn12\ which
allows to directly and independently determine the full set of
magnetic characteristics.

\Mn12\ was
pressed into  a pellet with a thickness of 1.39~mm and a diameter
of 10~mm. The high-frequency ESR experiments were performed by
means of a quasi-optical coherent-source technique
\cite{Volkov85}. The transmission spectra Tr($\nu$) of the pellet
were measured with a linearly polarized radiation (its magnetic
field denoted as ${\bf h}$), in the frequency range $\nu=8$ to
12~\cm\ at temperatures down to 1.8~K and in an external magnetic
field ${\bf H}$ up to 8~T applied perpendicular to the radiation
${\bf k}$-vector (Voigt geometry). One frequency scan over the
full range takes about  20 seconds.

Examples of the equilibrium  Tr($\nu$) spectra are
shown in Fig.~\ref{fig1}a. The absorption line at $\nu_0= 10$~\cm\
corresponds to the $|\pm 10\rangle\rightarrow |\pm 9\rangle$ CF
transitions at $\bf H$=0. It shifts to higher frequency
$\nu_{H}(H)\approx\nu_0+\gamma |H_z|$ ($\gamma = g\mu_B/h$,
$H_z=H\cos\theta_H$, $\theta_H$ is the angle between ${\bf H}$ and
the easy $C_4$ z-axis, $g\approx 2$) and changes its shape with
increasing ${\bf H}$, due to the inhomogeneous splitting of the
degenerate $|\pm 10\rangle$ and $|\pm 9\rangle$ states in the
polycrystalline \Mn12.

To study  {\em metastable} states of \Mn12\ we have investigated
the time dependence of the Tr($\nu$) spectra. Applying an external
field $H=+0.45$~T shifts the absorption line to the position
$\nu_{H}> \nu_0$ (Fig.~\ref{fig1}b); we then reverse the field (in
about a minute), which leads to the line shift to the position
$\nu_{L}\approx \nu_0-\gamma |H_z|<\nu_0$. Starting from this
point ($t=0$) the intensity of the $\nu_{L}$-mode decreases with
time while that of the $\nu_{H}$-mode recovers correspondingly.

The energy levels of \Mn12\
are determined by the effective Hamiltonian
\cite{Hill97,Barra97,Hennion97,Mukhin98}
\begin{equation}
H_{\rm eff} = DS_z^2 + BS_z^4 + C(S_+^4 + S_-^4)/2 - g\mu_B {\bf S}\cdot{\bf H} \label{eq1}
\end{equation}
where ${\bf S}$ is the effective spin of the ground $S=10$
multiplet, and D, B and C are the CF parameters. At small fields
the low energy levels are determined mainly by $H_z$: $E_{m} = D
m^2 + B m^4 - g\mu_B H_z m$, $m=0, \pm 1 \hdots \pm 10$ (inset of
Fig.1b). At low enough temperatures for $H_z>0$ only the
$|+10\rangle$ quantum state is populated  and  we observe only one
magnetic dipole transition $|+10\rangle$$\rightarrow$$|+9\rangle$.
After reversing the field ($H_z<0$) the state $|+10\rangle$
becomes meta\-stable and its population relaxes to the new ground
state $|-10\rangle$; the intensity of the transition
$|+10\rangle$$\rightarrow$$|+9\rangle$ (at $\nu_{L}<\nu_{0}$)
decreases while that for the transition
$|-10\rangle$$\rightarrow$$|-9\rangle$ (at $\nu_{H}>\nu_{0}$)
grows (Fig.1b). For our polycrystalline non-oriented sample a
random orientation of the easy-axis results in the population of both
$|+ 10\rangle$ and $|- 10\rangle$ CF states and in an
inhomogeneous broadening of the corresponding transitions by the
applied field.

New features in the relaxation phenomena appear at lower
temperatures, $T \lesssim 2$~K, as shown in Fig.~\ref{fig2}. The
relaxation occurs ``inhomogeneously''; i.e., it has a faster rate
near a particular frequency $\nu_-^{(1)}=9.57$~\cm\  where a peak
in the Tr($\nu$) spectrum appears accompanied by a minimum
developing at $\nu_+^{(1)}=10.47$~\cm. This new phenomenon can be
called {\em magnetic hole burning} in analogy to spectral hole
burning in dielectrics \cite{Galaup00}. We ascribe this fast
relaxation near the $\nu_{\pm}^{(1)}$ to a resonant quantum
tunneling in those crystallites whose orientation relative to
${\bf H}$ satisfy the level crossing condition
$h\nu_{\pm}^{(1)}(m) = h\nu_0 \pm g\mu_BH_z^{cros} = h\nu_0 \pm[-D
- B(m^2 +(m-1)^2)]$.

For a quantitative description of the observed relaxation the
effective magnetic permeability is calculated using (a) the
non-equilibrium populations of  the low CF states $|\pm 10\rangle$
and their evolution (relaxation) with time and (b) the
inhomogeneous broadening of the CF transitions in the magnetic
field due to the random orientation of the crystallites. Firstly,
we analyze a behavior for a single crystallite. Considering only
two
 $|\pm 10\rangle$$\rightarrow$$|\pm
9\rangle$ CF transitions and introducing a non-equilibrium
normalized populations $\rho_m(t)$ we generalize the ordinary
equilibrium permeability  \cite{Mukhin98} and obtain its
transverse components:
\begin{subequations}
\begin{eqnarray}
\mu_{xx,yy}(\nu,t) \equiv \mu_{\perp}(\nu,t) & =& 1+\Delta \mu^+(\nu,t) +\Delta \mu^-(\nu,t)\\
\mu_{xy,yx}(\nu,t) \equiv \pm \mu_{a}(\nu,t) &=& \pm i[\Delta
\mu^+(\nu,t) -\Delta \mu^-(\nu,t)]
\end{eqnarray}\label{eq4}
\end{subequations}
where
\begin{equation}
\Delta \mu^\pm(\nu,t) = \Delta\mu_0\left[\frac{\nu_0}{\nu_{\pm}}
(\rho_{\pm10} - \rho_{\pm9}) R^{\pm}(\nu)\right] .
\end{equation}
are the contributions of the $|\pm 10\rangle$$\rightarrow$$|\pm
9\rangle$ CF transitions to the  permeability with its static
value at $T=0$, $H=0$ being $\Delta \mu_0$; a Lorentzian lineshape
$R^{\pm}(\nu)$ is assumed for simplicity;
$h\nu_{\pm}=E_{\pm9} - E_{\pm10}=h\nu_0 \pm g \mu_BH_z$. The
equilibrium state is given by $\rho_m(t\rightarrow\infty) =
\rho_m^{\infty} = \exp\{-\beta E_m\}/Z$ with $Z=\sum\exp\{-\beta
E_m\}$, $\beta = 1/k_BT$, and $\sum\rho_m(t)=1$.

In the thermally assisted regime, the relaxation of the
populations $\rho_m(t)$ takes place via excited states and is
determined by the complex master equation for the populations of
all CF states \cite{Garanin97,Fort98,Leuenberger00,Pohjola00}. It
can be reduced to the simple relaxation equation $\Delta\dot{\rho}
= -({\Delta \rho - \Delta\rho^{\infty}})/{\tau}$ for $\Delta\rho =
\rho_{+10}- \rho_{-10}$ where $\Delta\rho^{\infty} =
\rho_{+10}^{\infty} - \rho_{-10}^{\infty}$ is the difference of
the equilibrium population after the field is reversed.  The
solution $\Delta\rho(t)=\Delta\rho^{\infty}[1-2\exp\{-t/\tau\}]$
for the initial condition $\Delta\rho(0) =\Delta\rho^0 =
-\Delta\rho^{\infty}$ leads to the population differences
\begin{eqnarray}
\Delta\rho_{\pm}(t)\equiv\rho_{\pm 10}-\rho_{\pm
9}=\Delta\rho_{\pm}^{\infty}+ \left(\Delta\rho_{\pm}^0 -
\Delta\rho_{\pm}^{\infty}\right) e^{-t/\tau} \label{eq7}
\end{eqnarray}
 which enter Eq.~(\ref{eq4}); here
$\Delta\rho_{\pm}^{\infty} = [\exp\{\-\beta \ E_{\pm 10}\} -
\exp\{\-\beta\ E_{\pm 9}\}]/Z$
  and
$\Delta\rho_{\pm}^{0} = [1 \mp \Delta \rho^{\infty}
(1+\exp\{-\beta h\nu_{\mp}\})] (1-\exp\{-\beta h
\nu_{\pm}\})/[2+\exp\{-\beta h\nu_{+}\} + \exp\{-\beta
h\nu_{-}\}]$ ,
where all quantities
refer to the reversed field. The derivation assumes thermal
equilibrium between $|\pm 10\rangle$ and $|\pm 9\rangle$ states
due to fast relaxation via phonons, implying $\rho_{\pm 9} =
\rho_{\pm 10} \exp\{-\beta h\nu_{\pm}\}$. For low temperatures
($\beta h\nu_{\pm}>> 1$), $\rho_{\pm 9}$ is negligible and
Eq.~(\ref{eq7}) simplifies to $\Delta\rho_{\pm}(t) \approx
\rho_{\pm 10}(t) \approx \rho_{\pm 10}^{\infty}+ \left(\rho_{\mp
10}^{0} - \rho_{\pm 10}^{\infty}\right) \exp\{-t/\tau\}$.

All previous consideration refers to a single crystal. For
polycrystalline  samples we have to take into account the
inhomogeneous broadening of the lines due to different splittings
of the doublets. By averaging the permeability [Eq.~(\ref{eq4})]
of the crystallites with respect to their orientations  we obtain
the effective permeability $\hat{\mu}^{\rm eff}(\nu,t)$ with the
non-zero components $\mu^{\rm eff}_{xx,yy}(\nu,t) = \langle
\mu_{\perp}(\nu,t)(1+\cos^2\theta_H)\rangle/2$, $\mu^{\rm
eff}_{xy,yx}(\nu,t) = \pm i \langle \mu_a(\nu,t) \cos \theta_H
\rangle$, and $\mu^{\rm eff}_{zz}(\nu,t) = \langle
\mu_{\perp}(\nu,t) \sin^2 \theta_H \rangle $ for the $\mathbf{H}$
oriented along the $z$-axis. Using $\hat{\mu}^{\rm eff}(\nu,t)$
and the Fresnel formulas for the transmission coefficient
\cite{DresselGruner} we fit the zero-field spectrum and obtain
$\nu_0 =10.02$~\cm, $\Gamma=0.18$~\cm, and $\Delta\mu_0=0.0107$,
in accordance with our previous results \cite{Mukhin98}. These
parameters also describe the spectra in the magnetic
field\cite{remark1} in all details, including the dramatic changes
of the lineshape (Fig.~\ref{fig1}a).

The relaxation phenomena observed in our Tr($\nu$) spectra at
various temperatures can be self-consistently described using the
theory of the magnetization relaxation in \Mn12\
\cite{Garanin97,Fort98,Leuenberger00,Pohjola00,Luis98} based on
phonon-assisted spin tunneling induced by forth order CF and
transverse magnetic field. To determine the relaxation rate
$\tau(T,H,\cos\theta_H)$ we have numerically diagonalized the
master equation for the density matrix $\dot{\rho}=W\rho$ in the
space of the ground $S=10$ multiplet, following the procedure
suggested by Ref.~\onlinecite{Pohjola00}. In calculations we used
the CF parameters of Eq.(1) $D=$-0.389 cm$^{-1}$,
$B=$-7.65$\times10^{-4}$ cm$^{-1}$, $|C| =5.0\times10^{-5}~{\rm
cm}^{-1}$ from \cite{Barra97,Hennion97,Mukhin98} and parameters of
the spin-phonon interaction determined via the main CF term
$DS_z^2$ \cite{Garanin97,Leuenberger00,Pohjola00}.

In Fig.~\ref{fig3} the angular dependence of the relaxation time
is plotted for two different temperatures and fields used in our
experiments
 (Figs.~\ref{fig1}b and \ref{fig2}).
The relaxation goes fastest for $\cos\theta_H =0$ because the
energy  levels on the both sides of the barrier coincide; a
situation, however, not observed in  Voigt geometry. For fields
exceeding the value of the first level crossing ($\sim 0.44$~T),
additional resonance minima arise with a fine structure related to
the slightly different resonance conditions for various exciting
levels due to the fourth order CF contribution B. As an example,
Fig.~\ref{fig1}b shows the relaxation observed in the Tr($\nu$)
spectra at $H=0.45$~T and $T=2.6$~K which are well described by
our model. We note that a weak resonance near $\cos\theta_H=0.97$
due to avoided level crossing of the CF states $m=4,-3$ (or
$m=3,-4$) does not exhibit in the spectra due to the averaging in
the polycrystalline sample. The relaxation rate strongly increases
with magnetic field because the effective barrier decreases with
both (longitudinal and transverse) components \cite{Barbara99b};
there are also significant contributions from the tunneling
processes which are enhanced because the tunneling splitting
increases with the transverse field.

These resonance effects are best seen at low temperatures, when
the thermal relaxation over a barrier is suppressed.
Fig.~\ref{fig3} shows a set of resonances in the relaxation rate
at $T=$1.96 K which correspond to pairs of states with an avoided
level crossing. Using this $\tau(\cos\theta_H)$, the time
evolution of the Tr($\nu$) spectra is calculated; as seen from
Fig.~\ref{fig2} the agreement is excellent. The main contribution
to the resonance relaxation comes from the thermally assisted
tunneling via (5,-4), (6,-5) and in less degree (7,-6) states (or
corresponding states with the opposite signs); they determine the
frequency  of the burned spectral hole. For these avoided level
crossings the resonance lineshape is determined by a truncated
Lorentzian \cite{Leuenberger00,Pohjola00} with the effective
linewidth determined by the tunneling splitting (0.23, 0.015, and
$5.0\times 10^{-4}~{\rm cm}^{-1}$, respectively) which is much
larger than the corresponding phonon linewidth (ranging from
$5\times10^{-6}$ to $3\times10^{-4}~{\rm cm}^{-1}$). The linewidth
($0.12 - 0.18$~\cm) of the zero-field-resonance absorption
$\nu_{0}$ exceeds the corresponding phonon linewidths
considerably. It is mainly determined by inhomogeneous broadening
and thus serves as a measure of the disorder effects
\cite{Mukhin01}, which could originate from CF and g-factor
distributions (D-strain and g-strain) and random magnetic dipolar
fields as was shown by ESR \cite{Park01}.

The problem of shape of the tunneling resonance and the effect of
inhomogeneous broadening on it are presently subject of
considerable interest and discussions
\cite{Barbara99b,Garanin97,Fort98,Leuenberger00,Pohjola00,Luis98,Friedman98,Prokof'ev00,Wernsdorfer99,Park01,Chudnovsky01,Amigo02},
many issues, however, remain open. One of them could be related to
the obvious discrepancy between the observed inhomogeneous
character of the zero-field modes broadening \cite{Mukhin01} and
Lorentzian shape of the tunneling resonances observed at
temperatures $T\gtrsim 2K$ \cite{Friedman98,Wernsdorfer99} and
described theoretically in the thermally assisted regime
\cite{Leuenberger00}. One may expect that a large inhomogeneous
broadening of the levels (i.e., larger than the tunneling
splitting) reduces the tunneling amplitudes and smears of the
resonances, resulting in suppression of the relaxation peaks and
in change of their Lorentzian shape \cite{Pohjola00} .
Suggesting that the main contribution to the line broadening comes
from the CF dispersion and assuming its Gaussian distribution
\cite{Park01}, we can evaluate the dispersion $\delta D=0.004$~\cm
of the main quadratic term $D$ in the Hamltonian (\ref{eq1})
\cite{remark2}. This yields a smearing of the tunneling level
differences $\delta (E_m-E_{-m+1}) = \delta D(2m-1) $, for
instance, of $0.035 - 0.05~{\rm cm}^{-1}$ for (5,-4), (6,-5), and
(7,-6) states. From a comparison with the calculated tunneling
splittings we can conclude that the inhomogeneous level
broadening does not effect the main resonance (5,-4), smears the
(6,-5), but suppresses the (7,-6) and higher order resonances. 
A numerical simulation including an additional averaging
with respect to the CF distribution confirms this conclusion.
Thus, if the tunneling splitting is sufficiently large for states
contributing to the thermally assisted tunneling,  the lineshape
of the corresponding resonance remains Lorentzian.

In summary, we utilized  a novel optical method of high frequency
magnetic spectrosopy to study the relaxation phenomena in \Mn12\
clusters, including the resonance quantum tunneling along with the
CF transitions and their lineshapes and linewidths. At elevated
temperatures only frequency-homogenous relaxation due to mainly
thermal activation is observed, whereas for $T \lesssim 2$~K the
thermal relaxation is significantly suppressed and resonant
quantum tunneling dominates; in this range the effect of {\em
magnetic hole burning } is observed. We have developed a
microscopic model which allows to quantitatively describe all
observed effects, taking into account the orientational
distribution of the relaxation time and  to determine the most
effective relaxation channels and to estimate the effect of CF
inhomogeneity on the resonant tunneling.

We thank the Prof.\ N. Karl and the Materials Lab for synthesizing
the \Mn12. We acknowledge support from the Deutsche
Forschungsgemeinschaft (DFG). This work was supported in part by
RFBR (No. 02-02-16597).

\begin{figure}
\caption{\label{fig1}Transmission spectra of polycrystalline
1.39~mm thick pellet of \Mn12\ measured in an external magnetic
field ${\bf H}\perp{\bf h}\perp{\bf k}$.
(a)~Shift of the absorption line with increasing magnetic field. The solid lines
represent Lorentzian fits with the distribution of the
crystallites orientation taken into account. (b)~Time evolution of
the absorption line after the magnetic field is reversed from
$+0.45$~T to $-0.45$~T. The lines represents the transmission
for the calculated angular distribution of the relaxation time
$\tau(\cos\theta_H)$ exhibit  in Fig.3.
The arrows indicate the positions of line at ${\bf H}$=0, $\pm 0.45~T$.
Inset: the energy levels of  a \Mn12\ single crystal  for a
magnetic field $\bf H\parallel C_4$; the arrows show the $|\pm
10\rangle \rightarrow|\pm 9\rangle$ transitions before and after
the field inversion.}
\end{figure}

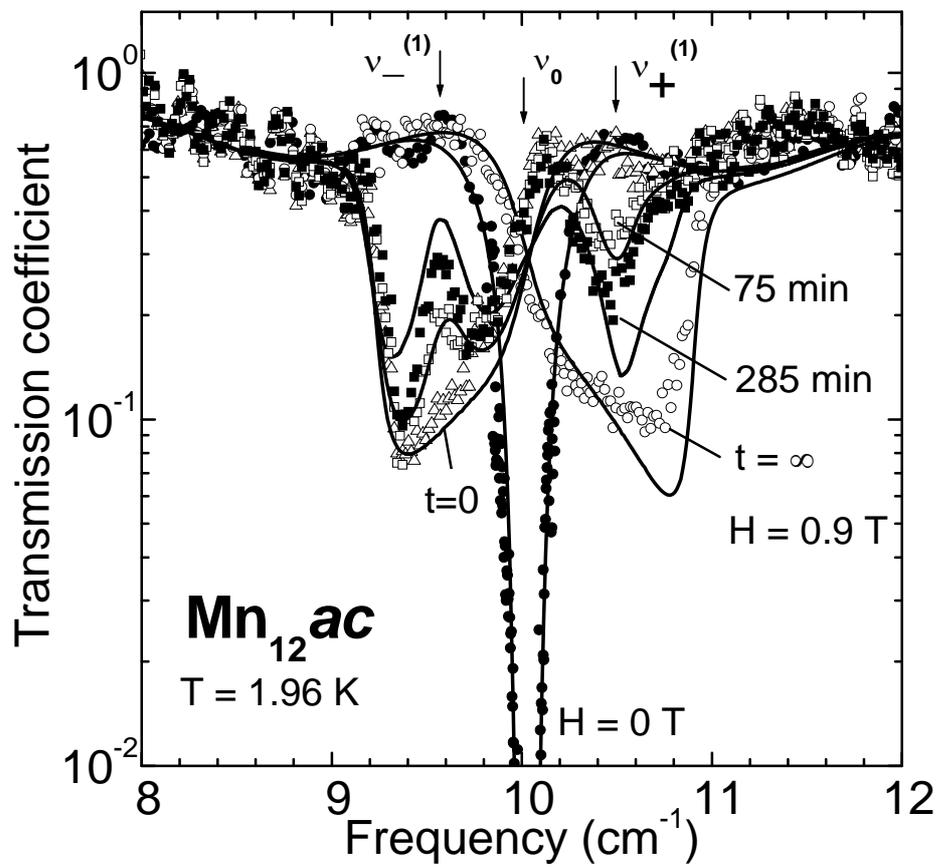
\begin{figure}
\caption{\label{fig2} Time evolution of the transmission spectra at $T=1.96$~K  after the magnetic field (${\bf H}\perp{\bf
h}\perp{\bf k}$) is reversed from $+0.9$~T to $-0.9$~T. The narrow absorption dip within
the broad feature is due to resonant quantum tunneling between pairs of coinciding energy levels.
The lines correspond to the transmission spectra for calculated distribution of the relaxation time  $\tau(\cos\theta_H)$ shown in Fig.3.
Also shown is the zero-field absorption line.}
\end{figure}

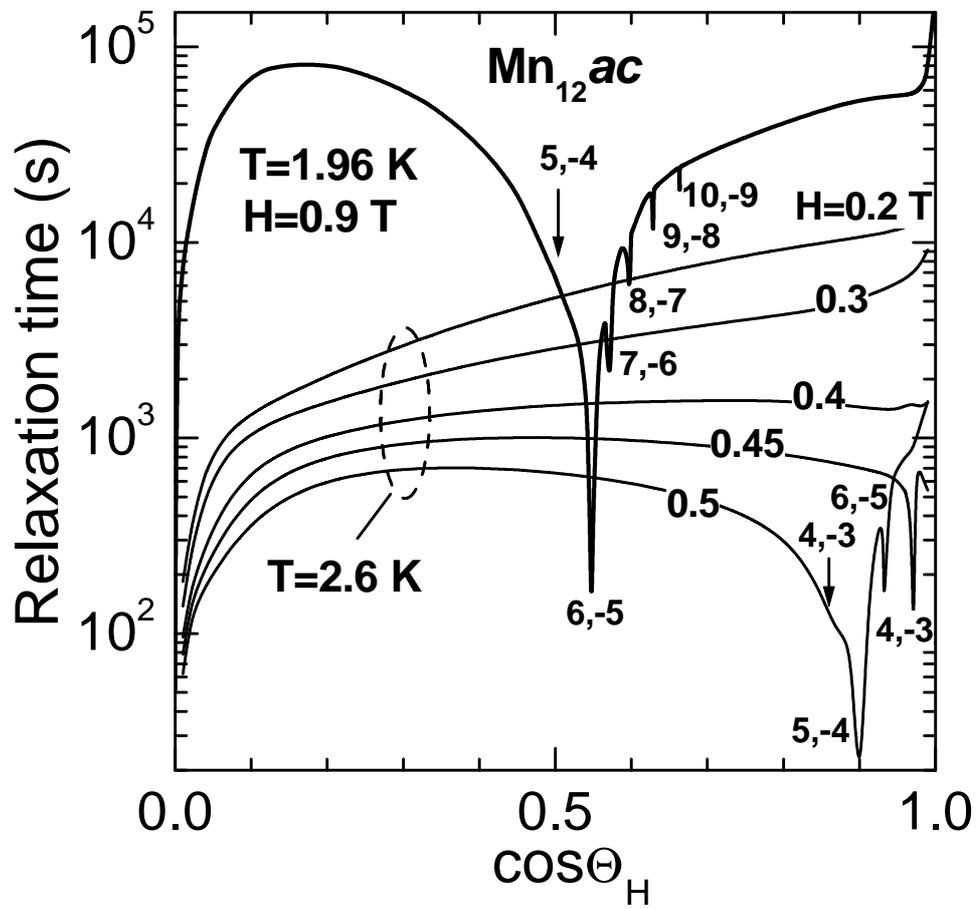
\begin{figure}
\caption{\label{fig3} Calculated distribution of the phonon-assisted
spin-tunneling relaxation time $\tau(\cos\theta_{H})$ as a function of
the angle between the magnetic field $\bf H$ and the $C_4$ axis of the
crystallites at $T=2.6$~K  and $T=1.96$~K
for certain values of the applied magnetic field. 
The numbers at the resonance minima indicate pairs of tunneling CF states.}
\end{figure}

\end{multicols}

\begin{thebibliography}{99}
\bibitem{remark1}Reversing the field we observe a weak asymmetry of the spectra with respect to the zero-field line
position $\nu_0$ which indicates a small internal dipolar field
$\lambda M \approx 0.04-0.05$~T; we have included it in our
calculations.
\bibitem{remark2}Imperfections of the crystal, such as
dislocations, can also promote spin tunneling by
inducing a transverse CF \cite{Chudnovsky01}; this
was recently observed by ESR in especially prepared \Mn12
samples \cite{Amigo02}.
\end{thebibliography}
\end{document}